\documentclass[prl,twocolumn,superscriptaddress,showpacs,%
amsmath,amssymb,floatfix]{revtex4}
\usepackage{graphicx}

\usepackage{epstopdf}

\newcommand{\Op}[1]{\boldsymbol{\mathsf{\hat{#1}}}}

\def\openone{\leavevmode\hbox{\small1\kern-3.3pt\normalsize1}}

\begin{document}

\title{Engineering an all-optical route to ultracold molecules in
  their vibronic ground state}

\author{Christiane P. Koch}
\email{ckoch@physik.fu-berlin.de}
\affiliation{Institut f\"ur Theoretische Physik,
Freie Universit\"at Berlin,
Arnimallee 14, 14195 Berlin, Germany}

\author{Robert Moszy\'nski}
\affiliation{Department of Chemistry, University of Warsaw,
  Pasteura 1, 02-093 Warsaw, Poland}

\date{\today}

\begin{abstract}
  We propose an improved photoassociation scheme to produce ultracold
  molecules in their vibronic ground state for the generic case where
  non-adiabatic effects facilitating transfer to deeply bound levels 
  are absent. 
  Formation of molecules is achieved by short laser pulses
  in a Raman-like pump-dump process where an additional near-infrared laser
  field couples the excited state to an  auxiliary state.  The
  coupling due to the additional field effectively changes 
  the shape of the excited state potential and allows for efficient
  population transfer to low-lying vibrational levels of the
  electronic ground state. Repetition of many pump-dump sequences
  together with collisional relaxation allows for accumulation of
  molecules in $v=0$.
\end{abstract}

\pacs{32.80.Qk,34.30.+h,33.80.Ps,42.50.Hz}
\maketitle

\paragraph*{Introduction}
Intense  interest in
ultracold molecular processes is generated by 
novel applications in ultracold chemistry
\cite{KremsIRPC05}, quantum information processing
\cite{MicheliNatPhys06}, or high-precision 
measurements \cite{ZelevinskyPRL08}. 
Ultracold molecules present themselves 
as ideal candidates also for coherent control: 
The utilization
of constructive and destructive interferences between different quantum
pathways in order to steer a process toward the desired target is not
hampered by thermal averaging.
Photoassociation (PA) provides a natural framework for merging the 
fields of  ultracold molecules and coherent control. It relies in
principle only on the presence of optical transitions:
Molecules are created by exciting
two colliding ultracold atoms to an electronically excited state with laser
light \cite{JonesRMP06}. In a few special cases the shape of the excited state potential
causes the probability 
amplitude to pile up at short distance, and molecules in the electronic ground state
can be formed by spontaneous or stimulated emission
\cite{ClaudePRL01}. 
PA, as well as photostabilization, can be optimized by a suitable
design of laser fields \cite{MyPRA06a,MyPRA04,PeerPRL07}. 
First experiments aimed at PA with short laser 
pulses 
\cite{SalzmannPRA06}
had to struggle, however, with difficulties due to the large spectral
bandwidth of femtosecond laser systems and the slow timescales of cold
collisions. A recent femtosecond pump-probe experiment could provide
evidence for  coherent formation of molecules \cite{Salzmann07}.
These excited state molecules have huge bond lengths, i.e. the
corresponding wavepackets reside at very large internuclear distance. In
order to dump them to the electronic ground state, possibly into a
single low-lying vibrational level, the molecules need to be brought
to short internuclear distance. Here we address the question how such
'$R$-transfer' can be achieved for generic excited states.

\begin{figure}[b]
  \centering
  \includegraphics[width=0.92\linewidth]{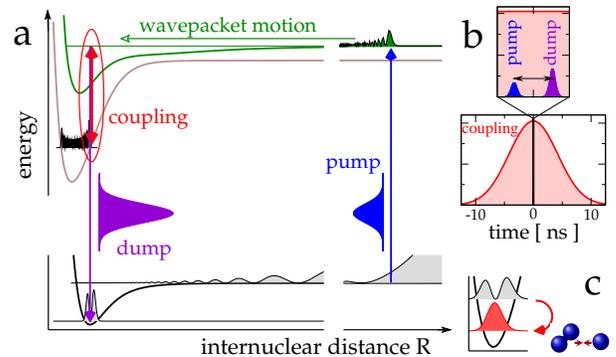}
  \caption{(color online) Pump-dump photoassociation scheme with a
    near-IR laser field providing resonant coupling
    between two excited states: potentials (a),  timing of the
    three fields (b) and collisional relaxation to $v=0$ (c).
  }
  \label{fig:schemegen}
\end{figure}

\paragraph*{Field-induced resonant coupling}
We adapt the coherent control concept of applying
a strong (near-) infrared laser field in order to
modify the excited state dynamics \cite{SussmanScience06}.
As shown in Fig.~\ref{fig:schemegen}a, the field 
\textit{induces} a coupling between two otherwise isolated excited
state potentials. In previous work, both experimental
\cite{SussmanScience06} 
and theoretical \cite{Gonzalez-VazquezCPL06},  
the infrared laser field was employed to \textit{suppress}
non-adiabatic transitions. 
Here, the two excited states 
are dressed by the coupling field, and the diabatic
picture is more appropriate:
The coupling field 
mimics resonant spin-orbit coupling, as it is found, e.g.,
 in the 
 $0_u^+$  states of heavy alkali
 dimers \cite{ClaudePRL01}. 
Resonant spin-orbit coupling leads to 
appreciable  binding energies ($E_b>1\,$cm$^{-1}$) of photoassociated
molecules in the electronic ground state
\cite{ClaudePRL01,HyewonMyPRA07}. Inducing the coupling by an external
field offers the advantage that the position of the potentials' crossing can be
tuned. This paves the way to ground state levels with much larger binding
energies, all the way down to $v=0$.

\paragraph*{Molecule formation}
We demonstrate that field-induced resonant coupling allows for
'$R$-transfer' in a generic excited state potential within a
two-color pump-dump scheme \cite{MyPRA06a,MyPRA06b}.
As illustrated in Fig.~\ref{fig:schemegen}a, molecules in the electronic 
ground state are formed by (i) a pump pulse exciting a wavepacket in the
electronically excited state at large internuclear distance,
(ii) wavepacket propagation toward shorter distances, where amplitude
gets 'trapped' due to the coupling with an auxiliary excited state, 
and (iii) a dump
pulse catching the wavepacket at short distance to transfer it
to the electronic ground state.
The coupling field amplitude is assumed to
be constant during the sequence of  pump and dump pulses.
Such a constant amplitude can be realized by a nanosecond
pulse, given that pump and dump pulses are of 
a few picoseconds full-width half-maximum
(FWHM) and that wavepacket motion takes  $50-100\,$ps, cf. Fig.~\ref{fig:schemegen}b.

\paragraph*{Accumulation of ground state molecules}
The repetition rate of short-pulse laser systems allows for collecting 
molecules over many identical pump-dump sequences provided that two
conditions are fulfilled. (1) The time between
two sequences is long enough for the system to equilibrate
to the same initial state. (2) Molecules transferred to $v=1$ by the
dump pulse must be removed from
$v=1$ before the next dump pulse arrives in order not to be
reexcited. A truly irreversible scheme is obtained if the molecules
\textit{decay} to $v=0$, e.g. due to collisions with atoms. 
While condition (1) is easily fulfilled for a kHz-repetition rate, the
timescale related to condition (2) is given by the inverse of the
collisional rate coefficient ($\sim10^{-10}\,$cm$^3/$s
\cite{StaanumPRL06})
times the density of atoms, i.e. $\tau$ varies between $1\,$s and
$10^{-4}$s from MOT to BEC densities.

\paragraph*{Model}
The Hamiltonian describing the 
situation depicted in Fig.~\ref{fig:schemegen}a reads  
\begin{equation}
  \label{eq:Hgeneric}
  \Op{H}_\mathrm{gen} =
  \begin{pmatrix}
   \Op{H}_{g}  & \Op{\mu}_1 \cdot E(t) & 0 \\[1ex]
   \Op{\mu}_1 \cdot E^*(t) & \Op{H}_{e} & \Op{\mu}_2 \cdot E(t) \\[1ex]
   0 & \Op{\mu}_2 \cdot E^*(t) & \Op{H}_{aux} \\
  \end{pmatrix}\,,
\end{equation}  
where $\Op{H}_i=\Op{T}+V_i(\Op{R})$ denotes the  vibrational
Hamiltonian of single channel $i$ ($i=g,e,aux$),
$\Op{\mu}_j$ the transition dipole moment, and $E(t)$ the
laser fields, $E(t)=E_{0,1} S_1(t) \cos(\omega_1 t) + E_{0,2} S_2(t)
\cos(\omega_2 t)$. The pump and dump pulses can be considered
separately (with $E_1(t)$ corresponding to either one of them) 
since wavepacket propagation in the excited state is slow and 
the time delay between pump and dump pulses correspondingly long.
In  the rotating-wave approximation (RWA), $E_1(t)$ couples
only to $\Op{\mu}_1$, and $E_2(t)$ only to $\Op{\mu}_2$.
The Hamiltonian is represented on a grid
using an adaptive grid step 
\cite{SlavaJCP99}, and the
time-dependent Schr\"odinger equation 
 is solved with the Chebychev
propagator.  

\paragraph*{Application to alkaline earth dimers}
The $^{40}$Ca$_2$ dimer is chosen as our prototype system.
The interest in ultracold
alkaline-earth and alkaline-earth like systems
such as ytterbium had been triggered by the quest for new optical
frequency standards.
Using continuous-wave lasers, PA was observed
for calcium and strontium as well as
for ytterbium
near both the $^1S_0- \,^1P_1$ atomic resonance
\cite{ZinnerPRL00,TakasuPRL04,NagelPRL05} and  
the $^1S_0- \,^3P_1$ intercombination line
\cite{CiuryloPRA04,TojoPRL06,ZelevinskyPRL06}. However,   
the formation of molecules in the electronic ground state has
not been reported to date.

\paragraph*{Choice of electronic states}
The pump-dump sequence is chosen to utilize a dipole-allowed
transition proceeding via 
the $B^1\Sigma_u^+$ excited state. The 
excited state lifetime  $\sim 5\,$ns does not incur losses in 
a coherent molecule formation scheme completed in
$\sim$100$\,$ps.  The $B^1\Sigma_u^+$ 
 state is well suited for PA \cite{ZinnerPRL00}: Its $1/R^3$
long-range behavior  gives
rise to large free-bound Franck-Condon factors close to the
dissociation limit. However, its vibrational wavefunctions, cf.
Fig.~\ref{fig:schemegen}a,  are 
not favorable to the formation of molecules in their electronic
ground state, and spontaneous or stimulated emission will simply
redissociate the molecules.
\begin{figure}[bt]
  \centering
  \includegraphics[width=0.92\linewidth]{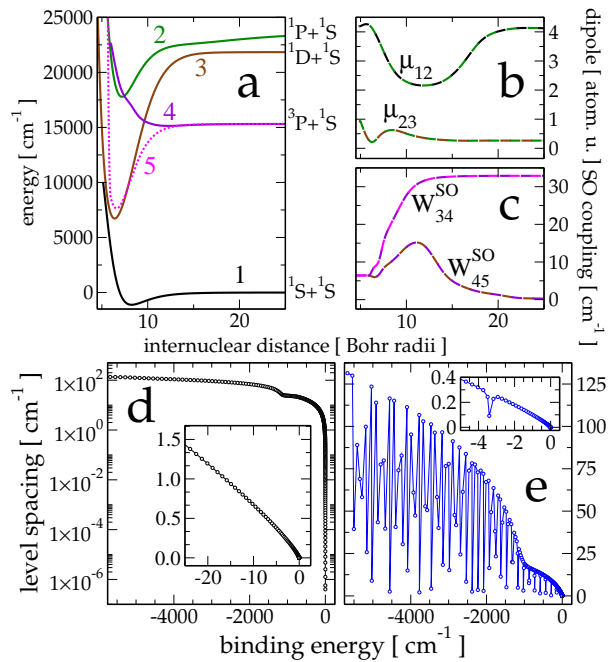}
  \caption{(color online) Minimal model for Ca$_2$ including all
    relevant physics: potential energy curves (a), transition
    dipole moments (b), and spin-orbit couplings (c) with labels
    corresponding as 1--$X^1\Sigma_g^+$,
    2--$B^1\Sigma_u^+$, 3--$(1)^1\Pi_g$, 4--$(1)^3\Sigma_g^+$,
    5--$(1)^3\Pi_g$. Spacings 
    of the vibrational levels of the $B^1\Sigma_u^+$ excited
    state (d) and of the $B^1\Sigma_u^+$ and $(1)^1\Pi_g$ states (e) dressed
    by a moderate near-IR laser field ($I=3.2\times
    10^9\,$W/cm$^2$, $\omega_L=881\,$nm). The red box in (d) indicates the
    range relevant for PA.
  }
  \label{fig:Ca2}
\end{figure}
Therefore the $B^1\Sigma_u^+$ excited state is coupled 
to an auxiliary state, chosen to be $(1)^1\Pi_g$. Of all states which
have a dipole-allowed transition to the $B^1\Sigma_u^+$ state, 
$(1)^1\Pi_g$ is closest in energy, with transition frequencies
corresponding to infrared (IR) and near-IR lasers. Such small transition
frequencies avoid undesirable multi-photon excitations which 
otherwise may be caused by the coupling field.
For $(1)^1\Pi_g$ as auxiliary state in
Eq.~(\ref{eq:Hgeneric}), a three-channel picture is not adequate
since the spin-orbit interaction
couples it at short range to the $(1)^3\Sigma_g^+$ state which in turn is coupled to the
$(1)^3\Pi_g$ state.
A minimal model adapting
Eq.~(\ref{eq:Hgeneric}) to Ca$_2$ 
comprises of the five channels shown in Fig.~\ref{fig:Ca2}a. 
The potential energy curves for calcium  are accurately known from  spectroscopy
\cite{DegenhardtPRA03} as well as from
state-of-the-art \textit{ab initio} calculations
\cite{BusseryPRA03}.
The potentials, transition dipole moment and spin-orbit coupling 
functions 
employed in the following calculations are gathered from Ref.
\cite{BusseryPRA03} and shown in Fig.~\ref{fig:Ca2} a-c.
In order to estimate how strongly the spin-orbit coupling perturbs
the levels of $^1\Pi_g$ state, the five-channel Hamiltonian
was diagonalized with $E_{1,2}(t)=0$.
While predissociation of some $^1\Pi_g$ levels is observed above
the $^3P+ \,^1S$ dissociation limit, 
below that
the effect of spin-orbit coupling is negligible. Utilizing only
$^1\Pi_g$ levels below the  $^3P+ \,^1S$ dissociation limit
corresponds to $\lambda_c<1200\,$nm for the coupling laser.
A model comprising of the three singlet channels is then sufficient.

\begin{figure}[bt]
  \centering
  \includegraphics[width=0.92\linewidth]{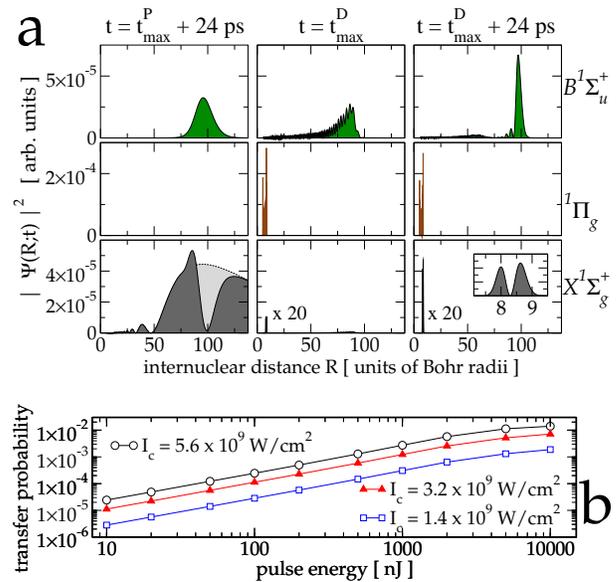}
  \caption{(color online) 
    Dynamics of the pump-dump photoassociation process under the influence of a
    moderately strong near-IR laser field ($I=3.2\times
    10^9\,$W/cm$^2$, $\lambda_c=881\,$nm): 
    The squared amplitude of the wavefunction
    in the three states is shown just after the pump pulse
    ($t=t^P_{max}+24\,$ps), at the maximum of the dump pulse
    ($t=t^D_{max}=t^P_{max}+t_{delay}$ with $t_{delay}=83\,$ps), and
    at the end of the whole process ($t=t^D_{max}+24\,$ps). The dashed
    line with light-grey filling in the lowest left panel
    corresponds to the initial scattering wavefunction. In the lowest
    middle and right panel only  population  which was transferred
    by the dump pulse to the electronic ground state is plotted.
    (b) Transfer probability from the excited states to $v=1$
    of the electronic ground state for different intensities of the
    near-IR coupling laser.  
  }
  \label{fig:dyn}
\end{figure}

\paragraph*{Wavepacket dynamics}
A sample pump-dump sequence is illustrated by wavepacket snapshots 
in Fig.~\ref{fig:dyn}a with time proceeding from left to right:
The initial scattering state (dashed line with light-grey filling in
Fig.~\ref{fig:dyn}a) describes two atoms
colliding in the $X^1\Sigma_g^+$ ground electronic state. A small
part of this state is excited by the 
 pump pulse (FWHM=10$\,$ps, $\Delta=-3.5\,$cm$^{-1}$,
$\mathcal{E}=0.3\,$nJ \footnote{
  Detunings $\Delta$ are taken with
  respect to the $^1P+^1S$ dissociation limit, pulse energies
  $\mathcal{E}$ are 
  calculated assuming a beam radius of 100$\,\mu$m.
}), forming
a wavepacket in the $B^1\Sigma_u^+$ excited state and leaving a hole
in the ground state wavefunction (left column of
Fig.~\ref{fig:dyn}a). Under the influence of the excited state
potential, this wavepacket moves toward shorter internuclear distances
where the coupling laser ($\lambda_c=881\,$nm, $I=3.2\times
10^9$W/cm$^2$) cycles population back and forth between the $B^1\Sigma_u^+$
excited state and the $^1\Pi_g$ auxiliary state. 
The time delay between pump and dump pulses is chosen such that the
peak of the dump pulse concurs with a maximum of population in the
auxiliary 
state (center column of Fig.~\ref{fig:dyn}a). The dump pulse
(FWHM=10$\,$ps, $\Delta=1016.2\,$cm$^{-1}$, $\mathcal{E}=1.0\,\mu$J) 
drives transitions from the excited 
state to the electronic ground 
state, populating $v=1$ (right column of
Fig.~\ref{fig:dyn}a). This population transfer is due to 
a \textit{dynamic} interplay of dump pulse and coupling field:
As the dump pulse depletes any excited state population 
within its Franck-Condon window, the coupling field refills
it, i.e. population  is channelled from the auxiliary through
the excited to the electronic ground state.
For stronger dump 
pulses, the depletion of the excited state 
occurs faster, and refilling this population shows a larger
effect.
The resulting dependence of the transfer probability on the dump pulse
energy is demonstrated in Fig.~\ref{fig:dyn}b. 

\paragraph*{Choice of the laser fields}
The wavelength of the coupling laser field, $\lambda_c$, dictates  the
internuclear distance $R_c$ at which the excited and auxiliary states
cross in a dressed-state picture, cf. Fig.~\ref{fig:schemegen}a. Since 
$R_c$ becomes the Franck-Condon point for the dump step,
$\lambda_c$ also determines the target 
vibrational  level in the electronic ground state.
To dump into  $v=1$ ($v=4$),
$\lambda_c$ needs to be 881$\,$nm (1080$\,$nm). 
The maximum value of $\lambda_c$ is given by the minimum difference potential of the
two states and amounts to roughly 5$\,\mu$m (corresponding to a target
level bound by $\sim 30\,$cm$^{-1}$). For wavelengths larger
than 1200$\,$nm, however, care has to be taken not to couple a predissociated
$^1\Pi_g$-level  to the $B^1\Sigma_u^+$ state. Depending on the
predissociation width, this could lead to a loss of population.
The intensity of the coupling laser field is determined by the
requirement that the coupling between the excited and the auxiliary state
potentials, $\Op{\mu}_2\langle E_2(t)\rangle$, be larger
than the vibrational level spacing in one of the potentials
(cf. Fig. 1 of Ref. \cite{Gonzalez-VazquezCPL06}). The level spacings
of the $B^1\Sigma_u^+$ state  are on
the order of 1$\,$cm$^{-1}$ for the range of
vibrational levels accessed by PA as shown in the inset of 
 Fig.~\ref{fig:Ca2}d.
The requirement on the coupling strength then
translates into peak intensities of the order $10^9\,$W/cm$^2$, assuming
a typical transition dipole moment of one atomic unit. 
  Such moderate intensities ensure that the coupling
  laser does not drive multi-photon transitions which could incur
  losses. Intensities at which multi-photon processes are to be expected
  can be estimated by comparing calculations 
  without invoking the RWA in Eq.~(\ref{eq:Hgeneric}) to those
  performed in the
  RWA. Increasing the intensity of the coupling laser reveals a
  breakdown of the RWA at $I\sim 1 \cdot 10^{10}\,$W/cm$^2$. 

For the pump-dump sequence,
pulses with transform-limited FWHM of a few picoseconds are best suited
with respect to both spectral bandwidth and timescales
\cite{JiriPRA00,ElianePRA04,MyPRA06a,MyPRA06b}.
Pulse energies of less than 1$\,$nJ are sufficient in the pump step 
to excite all population within the resonance window (cf. lower left
panel of Fig.~\ref{fig:dyn}a). Stronger pulses will not lead to
higher PA efficiencies. For the dump step, however, increasing the
pulse energy from 10$\,$nJ to 10$\,\mu$J yields better population
transfer to the target level, cf. Fig.~\ref{fig:dyn}b 
(such pulse energies can be produced by an amplifier). 
The detuning of the dump pulse is obtained in terms of 
the binding energies of the target level and  of the excited
state wavepacket, i.e. the pump detuning. The pump pulse detuning needs to
be chosen such that those excited state levels are populated which 
are indeed perturbed by the coupling laser.
The  position of the perturbed levels
depends rather sensitively on the coupling. Most likely it cannot be
theoretically predicted with sufficient accuracy. However, as
Fig.~\ref{fig:Ca2}e indicates,
the position of the perturbed levels and hence the pump pulse
detuning can be determined experimentally,  by 
performing spectroscopy on the dressed states. 
In analogy to resonant spin-orbit
coupling in alkali dimers \cite{HyewonMyPRA07}, the perturbed levels
are identified by peaks in the level spacings (or in the rotational
constants) as a function of binding energy, cf. Fig.~\ref{fig:Ca2}e.  
All other information required for the
implementation of the proposed scheme can reliably be obtained from
theoretical calculations.

\paragraph*{Efficiency}
An average over all thermally populated initial states
\cite{MyJPhysB06} yields the number of excited state molecules created
by one pump pulse. Assuming  the MOT conditions of Ref.
\cite{HemmerichPRA02}, $N_{mol}=12.5$ is obtained. 
With a dump transfer probability of $10^{-2}$,
cf. Fig.~\ref{fig:dyn}b, and a 10$\,$kHz-repetition rate, one molecule 
per ms is created in $v=1$. Collisional decay to $v=0$ within $1\,$ms
requires a density of $10^{13}\,$cm$^{-3}$. For lower densities, 
cavity-enhancement can be used to speed up the decay
\cite{MorigiPRL07}. 

\paragraph*{Conclusions}
Resonant coupling in two excited state potentials
can be mimicked by applying a near-IR laser field. This 
paves the way for Raman transitions deep into the well of the
electronic ground state. Necessary ingredients for 
such  a pump-dump scheme are 
(i) an excited state potential with $1/R^3$ long-range behavior and
with a dipole
allowed transition to the electronic ground state and (ii) an auxiliary
state with a dipole allowed transition to the excited state in the
(near-)IR. If the equilibrium distance of the auxiliary state is
smaller than that of the electronic ground state, the vibronic ground
state can be reached in a single pump-dump sequence. The required
intensities of the near-IR field are moderate and can easily be
produced for nanosecond pulses. 
Limiting factors are the low PA yield due to the small
number of atoms at sufficiently short range, and the slow collisional
decay. The PA yield could be improved e.g. by flux enhancement
\cite{GensemerPRL98}, and faster decay can be achieved via
cavity-enhancement \cite{MorigiPRL07}. 
Field-induced resonant coupling can be utilized to enhance
transition probabilities in any other Raman-like pulsed scheme such as
that of Ref.~\cite{PeerPRL07} and is therefore
expected to be useful well beyond
photoassociation. 

\begin{acknowledgments}
  \paragraph*{Acknowledgements}
  We are grateful to A. Osterwalder and R. Kosloff for their
  comments on the manuscript and to the 
  Deutsche Forschungsgemeinschaft (Emmy Noether grant) and 
  the Polish Ministry of Science and Higher Education (grant 1165/ESF/2007/03)
  for financial support. 
\end{acknowledgments}


\end{document}